# Microscopy is All You Need

Sergei V. Kalinin,[1,a] Rama Vasudevan,[2,b] Yongtao Liu,[2] Ayana Ghosh,[3] Kevin Roccapriore,[2] and Maxim Ziatdinov,[2,3,c]

[1] Department of Materials Science and Engineering, University of Tennessee, Knoxville

[2] Center for Nanophase Materials Sciences, and [3] Computational Sciences and Engineering Division, Oak Ridge National Laboratory, Oak Ridge, TN 37831

We pose that microscopy offers an ideal real-world experimental environment for the development and deployment of active Bayesian and reinforcement learning methods. Indeed, the tremendous progress achieved by machine learning (ML) and artificial intelligence over the last decade has been largely achieved via the utilization of static data sets, from the paradigmatic MNIST to the bespoke corpora of text and image data used to train large models such as GPT3, DALLE and others. However, it is now recognized that continuous, minute improvements to state-of-the-art do not necessarily translate to advances in real-world applications. We argue that a promising pathway for the development of ML methods is via the route of domain-specific deployable algorithms in areas such as electron and scanning probe microscopy and chemical imaging. This will benefit both fundamental physical studies and serve as a test bed for more complex autonomous systems such as robotics and manufacturing. Favorable environment characteristics of scanning and electron microscopy include low risk, extensive availability of domain-specific priors and rewards, relatively small effects of exogeneous variables, and often the presence of both upstream first principles as well as downstream learnable physical models for both statics and dynamics. Recent developments in programmable interfaces, edge computing, and access to APIs facilitating microscope control, all render the deployment of ML codes on operational microscopes straightforward. We discuss these considerations and hope that these arguments will lead to create novel set of development targets for the ML community by accelerating both real world ML applications and scientific progress.

[a] sergei2@utk.edu
[b] vasudevanrk@ornl.gov
[c] ziatdinovma@ornl.gov

The last decade has seen the meteoric growth of methods and applications of deep learning (DL). Starting with the seminal paper by Hinton, Krizhevsky, and Sutskever[1] that introduced deep learning for image recognition tasks trained with the ImageNet database by Fei-Fei Li[2], the field has developed almost exponentially, as evidenced by both numbers of papers, as well as recognition of the field in everyday society. The early overviews[3, 4] have laid the landscape for the applications of deep learning in a variety of areas, from medicine to computer vision to robotics. Over the last 2-3 years, we have witnessed an accelerating trend of adoption of deep learning in more experimental sciences, including chemical synthesis,[5] materials science and condensed matter physics,[6, 7] as well as multiple modalities of chemical and physical imaging.[8]

The rapid development in deep learning architectures for classical neural network architectures resulted in new models including different forms of convolutional networks,[9] introduction of graph networks (e.g., crystal graph convolutions),[10] and transformers,[11] by now ubiquitous for natural language and increasingly, vision applications. It has also stimulated the development of generative models such as variational autoencoders (VAE),[12-14] and generative adversarial networks (GANs)[15] and their variants that enable a broad range of generative and style transfer applications. Finally, advancements in deep reinforcement learning gave rise to such examples as muZero,[16] AlphaZero,[16, 17] AlphaFold,[18] RosettaFold,[19] and OpenAI5[20].

It is important to note that this progress has been based on multiple algorithmic and technical developments over the preceding decades. Neural networks were developed as an idea by McCulloch and Pitts in 1943.[21] Advances in computing would facilitate the invention of the first neural network *in-silico*, termed the ADALINE network,[22] which was used to filter audio signals. Multi-layer neural networks, studied first by Rosenblatt,[23] were explored in the 1970s,[24] but they were not significantly utilized until the backpropagation method was developed in 1986 by Hinton, Williams and Rumelhart[25] to enable efficient training. Reinforcement learning has been known for similarly long periods of time, going back to the works of Widrow in the late 60s,[26] with other pioneering work by Sutton.[26] The *q*-learning algorithm was developed in 1989 by Chris Watkins.[27] The long short-term memory (LSTM) network was introduced by Schmithuber in 1997,[28] and recurrent neural networks were studied for a decade before that seminal work. However, it is only over the last decade that the confluence of available computation capabilities and large data sets, emerging in the context of internet searches, social media, and open knowledge projects, that these methods became practically useful across multiple domains and applications.

In parallel with the development of algorithmic tools, the community has developed a range of static data sets, mainly driven by the need to define and benchmark the ML methods. In classical image analysis these include the paradigmatic MNIST hand written digits,[29] and multiple CIFAR data sets.[30] Similar data sets have been developed in domain areas, including theoretical quantum science, e.g. with the NTangled dataset,[31, 32] several examples in climate science,[33-35] as well as in materials science, for example the SuperMat dataset for superconducting materials,[36] or that of multi-principal element alloys.[37] Note that there is also a large number of datasets of simulated data in materials science, many of which are discussed in recent reviews.[38]

Much progress over this last decade has been fueled by the availability of these well-established datasets enabling benchmarking. While remaining a topic of considerable controversy, it is being argued currently that developing of the large 'foundation' models such as GPT and DALLE may be almost exhausted.[39] Indeed, these models clearly show outstanding performance for in-distribution tasks. However, they very often fail to generalize or yield a well-defined answer for a specific question (as anyone who played with DALLE[40] or Craiyon[41] have discovered). Unsurprisingly, the first applications of DALLE in science is cover art. Complementing these, somewhat less visible are problems of extreme training in costs that make these models available only to large research organizations (and very often trained models are proprietary). For example, the compute time for GPT-3 was 3650 petaflop-days,[42] which is about 34 days of continuous use of 1024 Tesla A100 GPUs. The energy for this training is about 190,000 kWh, which is about what 19 US homes use on average in an entire year.[43] Similarly, emerging is the issue of data exhaustion, i.e. models can be trained with full Wikipedia only once.

Complementary to it are by now multiple cautionary stories where DL models trained on static data sets are limited, and fail to generalize. Multiple examples of this problem abound, for example in self-driving cars, radiology and biological imaging.[44-49] It was the experience of the authors of this opinion piece that deep convolutional neural networks (DCNNs) trained on theoretical (i.e., simulated) data requires tuning for specific microscope settings,[50] but will fail to generalize for a different set of microscope parameters. While solutions based on more complex contrastive losses such as Siamese networks[51] and Barlow twins,[52] ensembling,[53] and invariant risk minimization[54] are being continuously developed, this problem is far from resolved.

A number of the leading scientists now argue that AI needs to move away from ever-increasing dataset and network sizes, to essentially 'do more with less'[55, 56] by using small data sets and physical priors. Similar arguments have been made in the physical science field,[57, 58] typically in cognizance of rich systems of prior knowledge and inferential biases and extremely small experimental budgets. This convergence between ML and physical fields is, in our opinion, a significant aspect of the last three years.

Traditionally, it is believed that machine learning methods emerge at the interface between classical computer science and established applications such as robotics, automated vehicles, and computer vision, are developed within these domains, and then become accepted in domain communities. However, for the domain experts it is immediately clear that adoption of the problems and challenges in the ML community often follows an "inverse" logic, with inherently complex and high-risk problems accruing most of the attention and investment, whereas relatively simple and well-defined problems are ignored or explored only in the domain context. At the same time, many domain problems offer highly simplified toy problems for ML. For example, in the table below, we establish the similarity between the challenges in the context of AI for self-driving vehicles and modern microscopies.

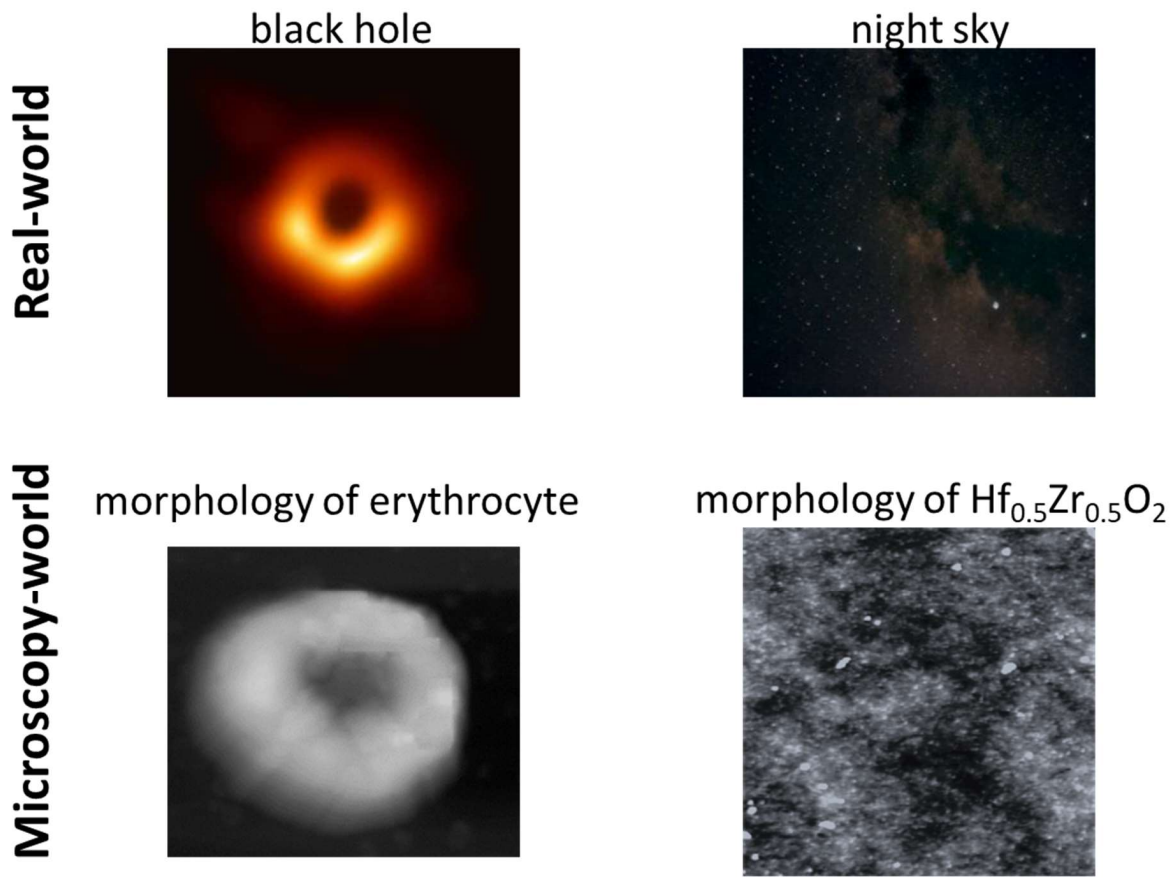

**Figure 1.** A comparison of real-world images and microscopy images shows the similarity of real world and microscopy world. The real-world image of night sky is generated by Craiyon[41]. The erythrocyte morphology image is adapted with permission from ref[59]. The HfZrO morphology image is adapted with permission from ref[60].

**Table 1. Microscopy is all you need (for now)**

| Autonomous driving | Electron Microscopy | Scanning Probe Microscopy |
|---|---|---|
| Camera glitch | Cosmic ray spark, electron gun change | Tip changes |
| Weird weather | Unexpected surface contaminations | Surface contamination |
| Camera out of focus | Microscope focus | Setpoint setting/sample drift |
| Optimize motor | Tune the microscope | Tune feedback |
| Conduct actions (turns, navigation) | Navigate to different parts of sample | Navigate to different parts of sample |

| Simulated drives for data collection | Simulated images from scattering codes | Simulated images from phase-field models or density functional theory |
| --- | --- | --- |
| Collision | E-beam induced sample damage | Tip crashes into a surface |
| N/A | Modify material with electron beam | Modify material with tip pressure or bias |

In this opinion, we argue that a promising new venue for the development of active learning methods capable of operating in real world situations are electron and scanning probe microscopy, both for intrinsic discovery value, and as simplified models for more complex systems. These methods are relatively low risk (compared to automated driving), are usually fairly closed systems minimizing but not excluding exogenous effects (unlike e.g. stock-market), come with the multitude of domain specific reward functions, allow for multiple levels at which prior knowledge can be incorporated, rely on the systems with well-defined but often unknown or partially known physical models, and allow for interventional and counterfactual studies. Most importantly, over the last years Python APIs and remote operation has become broadly available, allowing for the direct (via edge computing) and remote deployment of the codes on operational machines and their digital twins. Below, we expound on these considerations and (following a well-established ML meme) argue that electron and scanning probe microscopy is what the ML community needs (for now).

## 1. What are electron and scanning probe microscopy

When writing this opinion, we feel it is prudent to briefly introduce Scanning Transmission Electron Microscopy (STEM) and Scanning Probe Microscopy (SPM) to an ML community. STEM is an imaging technique based on the electron beam, as illustrated in Figure 2 (a).[61] The beam is emitted by the electron gun and subsequently focused and monochromated via a bespoke collection of electron lenses. Effectively the beam is compressed to sub-atomic width and formed byelectrons with a very narrow energy distribution. The STEM sample is a very thin membrane, typically below 50 nm thickness, i.e. ~1000 atoms. These samples can be fabricated via a variety of methods ranging from the native formation via crushing or delamination to highly sophisticated techniques such as focused ion beam (FIB) milling. Generally, the infrastructure for this sample fabrication is by now well established, and is available in academic, national labs, and industrial centers. Alternatively, paraphrasing Andrew Ng, don't worry about it.

Upon transmission through the sample, the electrons are collected via bright field (direct transmission) and dark field (electrons that were partially deflected) detectors. The signal intensity thus provides the information on the density of the solid. Perhaps the simplest analogy for STEM image formation mechanisms is shining the flashlight onto the construct and observing the shadows that transmitted light makes on the wall. Despite apparent simplicity, STEM allow direct visualization of atomic structures, with often spectacular images of atomic structures of metals, oxides, and semiconductors being available. In certain cases, the STEM images can be quantified, providing direct information on atomic spacing and bond length[62] – via direct images!

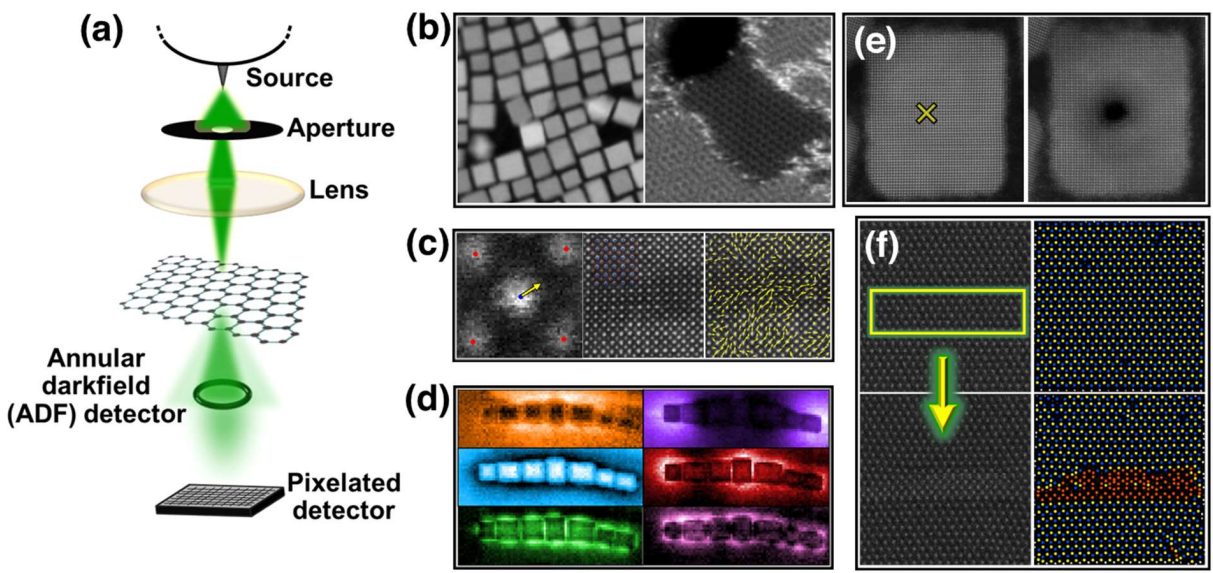

**Figure 2.** Principle of STEM as both a multimodal imaging tool and atomic manipulation tool. Schematic representation of electron path (**a**) as it interacts with a specimen and produces a variety of signals that may be detected, most standard of which is the annular dark field (ADF) signal. A magnetic prism (not shown) is typically at the end of the path that separates electrons based on energy lost (EELS) which is yet another signal. Examples of typical STEM imaging (**b**) where nanoparticles (left) down to single atoms in graphene (right) can be observed. Quantitative imaging example (**c**) by comparing ideal atom location with actual position, revealing slight displacements on the order of picometers of each cation (middle, blue dots) then visualized as a quiver plot (right) with implications in ferroelectric polarization and strain. Collection of electron energy loss spectrum (EELS) at each probe position in, for example, a chain of plasmonic nanoparticles (**d**) results in 3D data cube of which certain spectral components can be visualized as different spatially-resolved energy signals, represented here as different colors. In certain conditions, the electron beam can strongly interact with matter to the point of removing atoms from of the atomic lattice entirely (**e**) which allows to sculpt matter with atomic precision. Manipulation of matter at the atomic scale with single-atom specificity requires identification and classification of each atom followed by accurate positioning of electron beam onto desired sites, where (**f**) shows the formation of single vacancy lines (SVLs) in $MoS_2$ by specifically knocking out single sulfur atoms in a designated pattern. The images are reproduced with permission from John Wiley and Sons.

In addition to imaging, STEM allows a number of other interesting possibilities. The energy distribution of transmitted electrons can be measured (much like optical spectra) in a technique called Electron Energy Loss Spectroscopy (EELS). These EELS spectra can also be localized on the level of single atoms[63] and provide information on chemical and orbital states of individual atoms, energy excitations and quasiparticles, etc. The STEM is highly relevant to applications such as quantum computing (beam induced photon emission, quasiparticles and other excitations). Most intriguingly, the electron beam can be used for atomic assembly,[64, 65] to remove or move single atoms[66, 67] and assemble atomic structures.[65, 68] For the time being, with few exceptions[69] the atomic manipulation was enabled with purely human operator control.

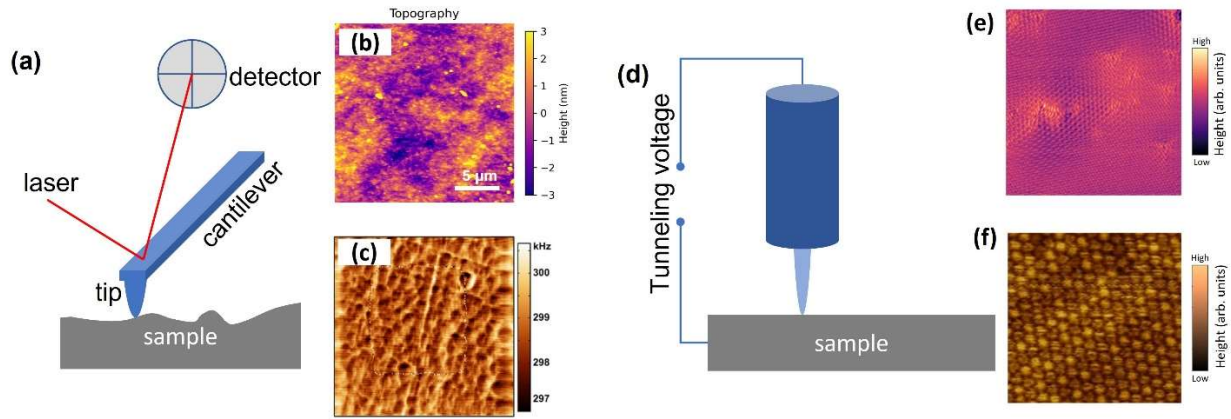

**Figure 3.** Overview of the scanning probe microscopy (SPM) operation, which includes force-SPM and scanning tunnelling microscopy (STM), (a) schematic of the force-SPM setup, (b), force-SPM topography image, adapted with permission from ref[60], (c), force-SPM resonance frequency image, adapted with permission from ref[70]. (d) schematic of STM setup, (e, f), STM images of atomic defects in graphene (e) and molecular self-assembly (f).

The second type of microscopy we discuss here is the Scanning Probe Microscope (SPM). SPM is based on the interaction of the sharp mechanical probe, or tip, with the surface. In force-based SPMs, the probe is fabricated on the end of a silicon cantilever and the interaction between the probe and the surface leads to the deflection of the cantilever (Figure 3). This deflection is then detected via either an interferometer or a laser beam deflection system.[71] Despite simplicity, the sensitivity of this method is such that the forces on the level of interatomic ones can be detected, hence the name atomic force microscopy.[72] The simplest but apt analogy to SPM is touching the surface with the finger (with eyes closed).

Depending on the specific configuration and operation mode, force-based SPMs can be made sensitive to a broad range of forces including magnetic, electric, and Van-der-Waals, allowing visualization of phenomena such as magnetic domains,[73, 74] electronic and ionic transport,[75-77] mechanical properties of the surface,[78] ferroelectric domains,[79] and many others. SPM can be implemented on active devices, exploring phenomena induced by lateral stimuli.[76, 77, 80, 81] The SPM tip can be further used to manipulate matter, from local nanooxidation and lithography[82] and writing ferroelectric domains[83] to mechanical manipulation, moving particle and nanowires around.[84-86] In fact, multiple versions of haptic interfaces for SPM have been developed over time providing a human interface to the nanoworld.[87-89]

A special (and first invented) class of SPMs is Scanning Tunneling Microscopy (STM).[90-92] STM uses simple metallic probe and detects the tunneling current between the probe and conductive surface. While generally requiring a ultra-high vacuum environment for operation,

STM is capable of visualizing surfaces on the atomic level and manipulating matter atom by atom. In fact, the letters "I", "B", and "M" written in xenon atoms on the copper surface by Don Eigler at IBM[93] was the initial demonstration of atomic manipulation, and gave impetus to the nanotechnology revolution. Now, manipulation of phosphorus atoms on silicon surfaces is broadly considered as one of the possible pathways to solid-state qubits[94, 95] for quantum computing.

Common for both SPM and STEM is the basic principle of a mechanical (SPM, STM) or electron beam probe that can be scanned across the material surface and detect the local state (imaging mode), perform more detailed measurements (spectroscopy), or modify the material state. The probe parameters and trajectory can be controlled by the microscope electronics. In this manner, we consider these to be systems that allow a limited set of operations and controls on the local level, and provide a convenient model system for deployment of machine learning, active learning and reinforcement learning algorithms. Below, we discuss these considerations further.

## 2. Why microscopy for machine learning

Electron and scanning probe microscopy are, by now, well-established fields with broad applications within physics, chemistry and materials science. Below, we discuss aspects of these field that, in our opinion, makes them ideal model systems for active learning methods and hence the ML community. In all cases, we introduce domain-specific considerations and illustrate the connection to the ML.

### a. Language and control

The vast majority of STEM and SPM studies have been performed with commercial tools, often representing black box systems with fairly narrow user interfaces and (for STEMs) service contracts. While customizable controllers have been available (e.g. *Nanonis*) they used to represent a small part of the field. Correspondingly, deployment of custom developed spectroscopies and machine learning tools required teams comprising microscopy, physics, and electrical engineering expertise and have been the exception rather than the rule.[96-98]

The situation started to change rapidly over the last several years, when a number of microscope manufacturers started to introduce Python programmable interfaces for commercial tools. A partial list includes SWIFT for NION microscopes,[99] PyJEM for JEOL systems[100] and even more open source efforts in optical microscopy that are significantly hardware agnostic, such as pycroManager.[101] For Scanning Probe Microscopy, this includes Python interfaces e.g., the NanoSurf atomic force microscopes,[102] and home-grown efforts at multiple research centers worldwide[103, 104] for control of low-level microscopy functions on a range of instruments, all via python commands.[105] Correspondingly, the operator has significant programmable control over the microscope operation, and thus correspondingly, access to data streams. This makes it significantly easier to deploy ML codes on these data streams to affect microscope operations, thus greatly lowering the activation barrier for deployment.

Additionally, the ability to 'script' entire experiments and run them on multiple instruments is a significant shift not only for the types of experiments that are then possible (due to automation), but also in that it changes the user paradigm to one of remote operation, in tune with the recently developed 'cloud labs' for chemical synthesis.[106, 107] This parallels the deployment of online quantum computers that allow users to run their own codes by simply uploading them.[108-110] Remotely operated and automated microscopy instruments offer the same tantalizing possibility, to test not only new experimental workflows, but to benchmark machine learning algorithms across a range of microscopy platforms.

The last decade has also seen significant developments in terms of data models and associated infrastructure for the microscopy community. Although there exist many formats, on almost all occasions typically there exist ways to convert between multiple formats and read data and associated meta-data into python, regardless of which instrument the data was acquired on. For scanning probe microscopy, virtually every vendor produces their own data format, but programs such as WsXM[111] and Gwyddion[112] have long been capable of reading of data from a wide variety of SPM formats. Tools such as pySPM[113] and Pycroscopy SciFiReaders[114] have significant collections of 'translators', which build on community tools to convert data into standardized data models, such as the spectral imaging dataset.[115] Regardless of the choice of final format, the codebase exists to convert most existing datasets to a common one for generating curated ML datasets.

A similar situation exists in the electron microscopy, with translators existing for virtually all of the commonly used microscopes, with packages such as HyperSpy[116] and openNCEM[117] providing the necessary bridges to ensure interoperability. In contrast with most scanning probe work, it is interesting to note that the idea of parallel computation and workflows is more established within the TEM community through efforts such as LiberTEM[118], primarily due to the nature of the reconstruction algorithms that are compute intensive. Note however, that this is changing rapidly within scanning probe microscopy in the past two years, and workflow solutions more specific to SPM can be expected to follow suit in due course.

Finally, we note that a significant portion of development on microscopy platforms occurs at scientific user facilities, in both the United States[119] and across other centers around the globe. This dynamic ensures continuous development that is not limited to groups led by individual PIs at universities, and furthermore, encourages a more broad-based effort, given the needs of the user center are to ensure new techniques and analysis methods are available to the user community with minimal barriers. Overall, we are not yet at the stage when the codes from GitHub can be deployed to run on selected microscope as a "physical RL gym" from the ML side. However, from the domain side it is already possible to run specific code as a plug-in to the microscope.

**b. Risk**

One significant consideration in real world applications of ML is that of risk. Perhaps the most recognized case for it is autonomous driving, where each accident is instantly popularized. However, this is also well recognized in areas such as ML fairness and explainability, with the algorithm trained on historical data often making predictions and recommendations inconsistent with current social norms.[120]

Comparatively, for microscopy the risk is relatively small. It is well understood that scientific research is prone to errors, and knowledge emerges as the result of multiple trial and error cycles, often including community feedback. Given this established paradigm, any results derived from the automated workflows will be verified by conventional methods for an extended time, and the impact of potentially incorrect findings can be expected to be comparable to that derived during normal research processes.

The second component is the risk to the instrument. While crashed kernels in the notebook can be restarted, the crash in a microscope often necessitates the change of the probe. For ambient SPMs this is of the order of ten minutes and process can be automated. For UHV systems it can be several days (and hence STM community is highly conservative). Beyond the probe, the damage for the mechanical parts of the systems may be considerable. Practically however, these risks can (and are) heavily mitigated by constraints in allowable operations. Following on the concepts and ideas from the operation of cloud compute infrastructure and automated labs, the allowed actions in automated microscopy can be chosen to be within safety tolerances to ensure equipment safety. Similarly, de risking is possible via the introduction of the Digital twins, meaning the *in-silico* models of the microscope that either emulate operation via ground-truth data, or allow modelling of the mechanical and electronic equivalent of the microscopes.[121]

Note that the advantage of the microscope as a model system for active learning is the low risk combined with the large variability of microscope and sample environments, ideal for developing ML solutions stable with respect to large out of distribution effects. Once developed, these can be potentially transferred to higher-risk problems, e.g. robotics or autonomous drones.

**c. Exogenous variables and causal models**

A very important consideration for any real-world ML algorithm is the number and character of exogenous variables, i.e., external (to the system) stochastic parameters. For example, the success of AlphaGo[17, 122] or any other reinforcement learning algorithm for computer games is predicated on known rules and lack of external variables (e.g., the number of squares on the board will never change). In other words, while the game contains the element of randomness, the general rules are constant, and the algorithm builds the approximations via the Q-tables or continuous policies.

Comparatively, the self-driving car or stock-trading algorithm has to deal with influences outside of the system – from pedestrians coming from around the corner to political events. Correspondingly, long-term prediction or even models of the environment for long time are difficult to construct. Clearly, the absence of exogenous variables is an oversimplification of the real-world applications (the world is not a chess game), while too many limits how far we can predict (it is not a fully random process either).

From this perspective, microscopy offers an intriguing range of possibilities from almost deterministic to highly stochastic systems and from full to partial knowledge of the system, all in the presence of strong physical priors. For example, STEM imaging of bulk materials visualizes the atomic columns averaged over the beam direction. Imaging with focal and tilt series, while more complex, provide information on the structural variability in the beam direction.[123-127] Atomically resolved images of 2D materials such as graphene allow all static atoms to be visualized in the image plane, but the observed phenomena may also be affected by the presence of (invisible) moving atoms that can participate in chemical reactions and atomic reconfigurations. At the same time, at this length scale the objects that we observe in the microscope can only be atoms, and atoms have a well-defined geometry. Correspondingly, ML analysis of such data allows for very strong inferential biases.

For SPM microscopy these considerations are more complex and can be less well defined depending on the specific material system. For example, UHV STM relies on atomically clean surfaces to minimize unwanted noises and chemical contamination (i.e., each stray atom will be seen as a "mound" on surface). Techniques such Piezoresponse Force Microscopy[128] allow almost complete decoupling of information on different aspects of materials functionality, e.g. topography, elastic properties, and ferroelectric domain structures. At the same time, Magnetic Force Microscopy[74] will be sensitive to both magnetic and electric interactions, whereas phase imaging mixes all interactions including magnetic, electrostatic, adhesion, and elastic into one information channel.

Yet, compared to largely uncontrollable interactions in the real world, microscopy offers an opportunity to explore correlative and causal links in technologically relevant systems under (mostly) controlled environments, starting from relatively simple systems with largely known physics (e.g., one-element-based 2D materials) and progressing towards more complex systems demonstrating yet-to-be understood physical behaviors (e.g., cuprate superconductors).

For example, scanning probe microscopy and spectroscopy allow collecting data on both atomic positions and physical behaviors of interest (encoded in spectra) over the same sample region.[129, 130] The local atomic distortions and classical and quantum phenomena are assumed to be correlated and their relationship is considered to be parsimonious, that is, ultimately explained by a small number of (latent) mechanisms. A conventional approach to the analysis of such data is to apply a blind unmixing to the hyperspectral data and do a linear correlation analysis of the abundance maps of the derived sources with structural descriptors at each unit cell.[131] It would be interesting instead to utilize deep latent variable models trained in end-to-end fashion (with structural images as input and spectroscopic curves, full or scalarized, as outputs) to disentangle different physical mechanisms and establish how changes in specific geometric features influence a property of interest (for example, local $T_c$). Furthermore, in complex systems with competing interactions and chemical disorder, it will be critical to distinguish between correlative and casual links, offering a playground for a newly emerging class of casual representation learning techniques.

### d. Interventions and counterfactuals

Electron and scanning probe microscopies are often perceived as a purely observational areas, much like astronomy. However, familiarity with the field suggest that this is generally not the case – both probes and beams can modify the surface, but these behaviors are often perceived as deleterious. For example, while a high energy electron beam can easily modify the sample, this is traditionally perceived as beam damage. In fact, much of the evolution of modern STEMs, first towards high voltages and then towards low voltage aberration corrected machines was driven by the need to balance resolution and beam damage. Similarly, in scanning probe microscopy the damage to the surface and especially the probe is generally unwanted.

However, for many materials classes the evolution of both SPM and electron microscopy techniques has facilitated control over the modifications inflicted on the sample. . Atomic manipulation in STM[93] and STEM[66, 68, 132] is one example of this. Here, we create atomic configurations that are absent in the original material, and explore their functionality via imaging and spectroscopy (Figure **4**(b)). Similarly, in piezoresponse force microscopy we can create new domain structures and explore their polarization dynamics.[79, 133]

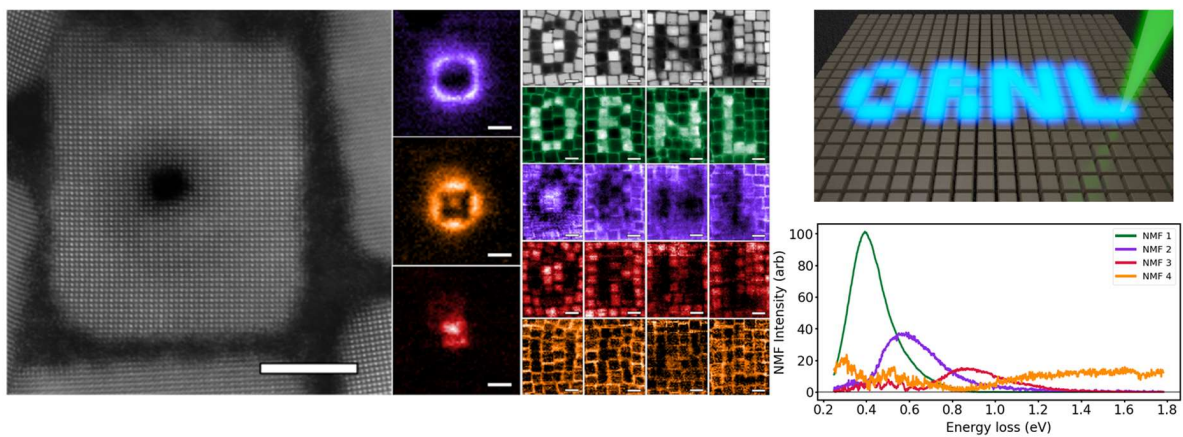

**Figure 4.** Electron beam sculpting of matter and its connection to its nanophotonic properties. Specific energetic features can be placed by the electron beam by sculpting nanoparticles into geometric shapes that support a desired nanophotonic property. Reprinted with permission from John Wiley and Sons.

From a machine learning perspective, this opens a pathway for exploring interventional and counterfactual questions implemented on an experimental platform. For example, a system of plasmonic nanoparticles contains multiple local geometries that can be extracted from the experimental data and correlated with the associated plasmonic spectra using im2spec[134, 135] or parallel VAE approaches.[136, 137] These methods allow us to predict the plasmonic properties for the extant geometries. However, the electron beam can be used to modify the system in a broad variety of scenarios, including those that remain in distribution compared to original data (e.g. remove the particle and create a hole similar to the ones that already exist), or create geometries that were not present in the original data sets (e.g. hole within the particle). Combined with the reasonably well-understood physics of these systems, this opens a very broad set of opportunities to implement causal active learning methods and explore interventional and counterfactual strategies.

**e. Prior knowledge and active learning**

Most ML methods are correlative and generally take only limited advantage of the prior knowledge, typically implicitly in the form of output or the loss function. As such, they capture the structure of the target distribution, but generally impose no limitations on the properties of the object. However, physical sciences take a very different approach where past knowledge plays the preponderant role, and the experiment is often formulated to falsify the specific hypothesis formulated based on prior domain specific and general knowledge.

Recently, significant progress has been achieved in developing physics-informed neural networks.[138] In these, the network structure can be such that outputs directly satisfy certain invariances or equations of motion. Alternatively, the known factors of variability such as rotations, translations, or shear can be incorporated as defined latent variables in architectures such as variational autoencoders.[139-141]

From the perspective of physics-informed machine learning, STEM and SPM offer a broad range of opportunities both on the instrument and materials side. From the instrument perspective, the SPM cantilever oscillations closely follow Euler beam equations, providing a strong inferential bias. In electron microscopy, the image formation mechanisms are complex, but the physics of the multiple scattering can be naturally incorporated into the neural network structure. An even broader range of opportunities emerge on the materials side, where materials systems chosen for detailed study usually come with known composition, preparation history, and hence well-defined physical priors in structure and range of possible phenomena.

The important aspect for ML applications is that the experimental setup in SPM is often selected to narrow down the scope of relevant knowledge. For example, when exploring ferroelectric domains,[142] the relevant information with a good degree of approximation is the domain structure tied to the corresponding free energy functional and the preparation method. At the same time, a second level of parameters with be the surface preparation and surface chemistry conditions.[143-145] Often the potential contributions of certain level of description can be estimated from straightforward physics-based estimates, with the remaining variables contributing to exogenous factors with some known level of significance.

**f. Physical models**

The characteristic aspect of many physical systems is the simplicity of the laws that govern their behavior, as exemplified by the fundamental gravitational laws, Maxwell equations, or quantum mechanics. This local simplicity however gives rise to very complex behaviors of real systems, from real material properties to the emergent behaviors in systems with collective interactions.[146] The important aspects of these laws is that generally they are expected to be universal, e.g. the force fields between two carbon atoms do not depend on where in the universe these atoms are. Correspondingly, it can be argued the high-resolution imaging studies can be aimed at discovery and identification of these laws.

The significant progress toward these applications of machine learning has been accomplished in astronomy. It has been demonstrated by Ho, Cranmer, and Battaglia that the symbolic laws of gravitation can be derived from observations of planetary motion.[147] Similarly, Tegmark and coauthors have demonstrated that equation of motions can be derived from video data.[148]

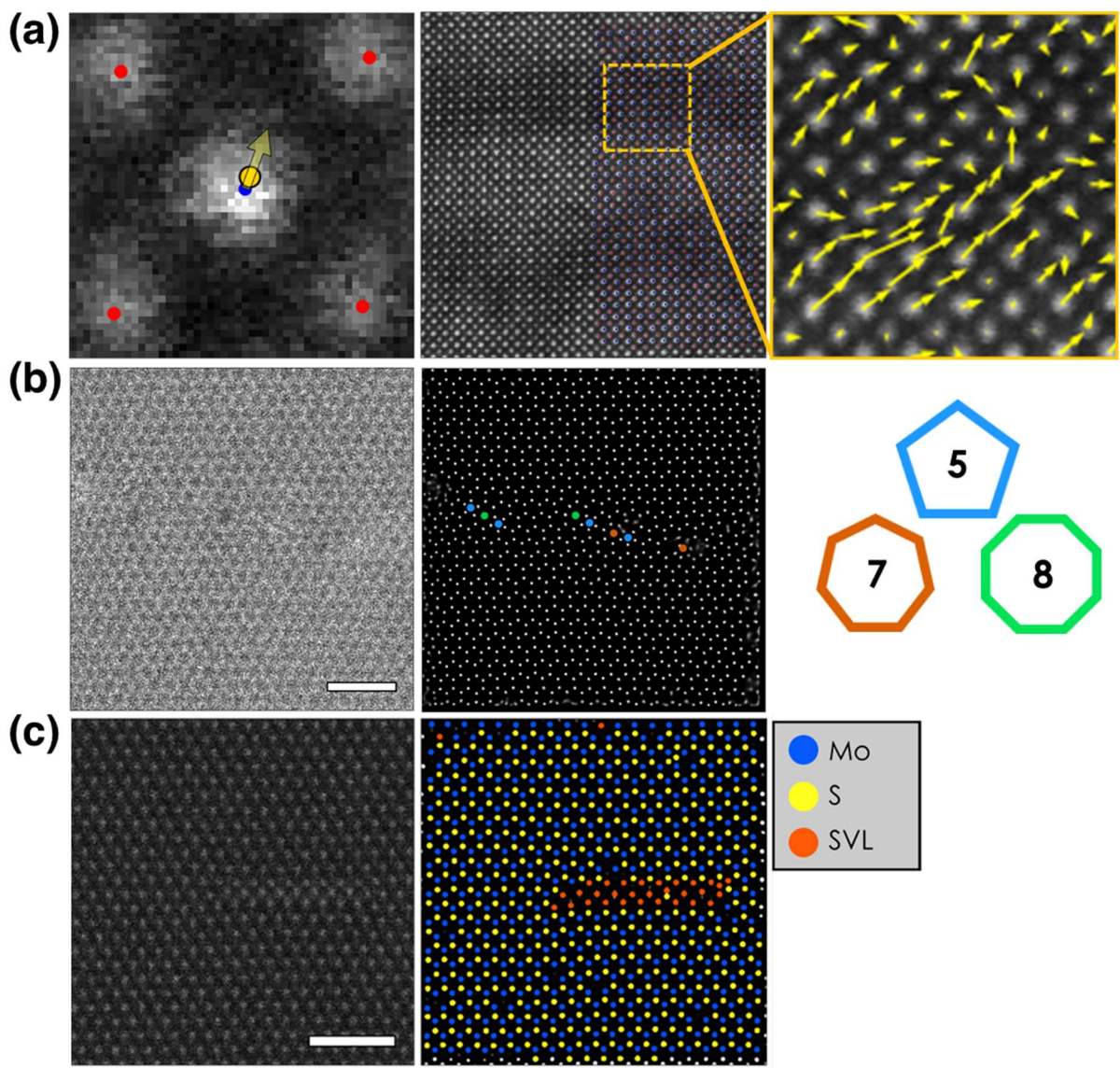

**Figure 5.** Feature and property extraction with use of atomic coordinates. Polarization and strain fields may be calculated once atom columns are properly located and classified, for example a ferroelectric material shown in (**a**) has its cations (bright atoms, blue) distinguished from the rest of the lattice (dim atoms, red). The deviation of the cation from its ideal position yields a displacement vector which can be interpreted as a built-in polarization field. Atomic classification in other materials allows to identify subtle features at the single defect level, such as Stone-Wales

defects or other topological ring defects in single layer graphene in (**b**), or sulfur vacancy lines in MoS$_2$ in (**c**).

From this perspective, microscopy data opens virtually unlimited opportunities to explore physical models. Provided only the position of atoms, it is possible to extract a number material properties. For example, built-in polarization fields and ferroelectric domains can be identified by – i.e., the electron beam in the STEM is sensitive enough to resolve picometer displacements that have origins in polarization.[149] Figure 5(**a**) demonstrates how the determined atomic coordinates in a ferroelectric superlattice are used to compute polarization fields – simply, the deviation of a cation (bright atom) from its ideal lattice position gives the polarization vector at that site. Similarly, it is known that strain can in turn have an impact on electronic properties[150], but it turns out only a small fraction of atoms deviate from their nominal lattice location, for example, at the interface of a core-shell nanoparticle. Hence, we can begin to understand the underlying reason why a material exhibits a certain response – it may be intimately tied to a very localized strain or built-in polarization field.

Another subtle feature that can be detected via atomic coordinates is the presence of defects, which can exist in the form of vacancies, impurity atoms, or practically anything that is not part of the pristine crystal. Many times, these can be extremely difficult for the human operator to notice (especially in the case of a single atomic defect) during experimental operation, therefore a rapid means for detecting and classifying atomic coordinates is of critical importance. In Figure 5(**b**), a single layer of graphene is shown, where the atomic coordinates given by a trained deep ensemble neural network[151, 152] are shown. Graph analysis is performed given the coordinates, from which vacancy defect structures like a Stone-Wales[153] (or, 5-7 ring) defect strongly stand out. As a separate example, in Figure 5(**c**), HAADF-STEM image of MoS$_2$ is shown also with its detected coordinates shown, where completely different defect structures can be identified and clustered based on, for instance, nearest neighbor distances or relative intensities. In the systems where the electron beam can create atomic defects, each image contains information on the position of multiple atoms in the system in the quasi-equilibrium state. The possible configurations are ultimately determined by the force fields between the atoms. Hence, observations of the atomic configurations allow (in principle) probabilistic reconstruction of the force fields between the atoms,

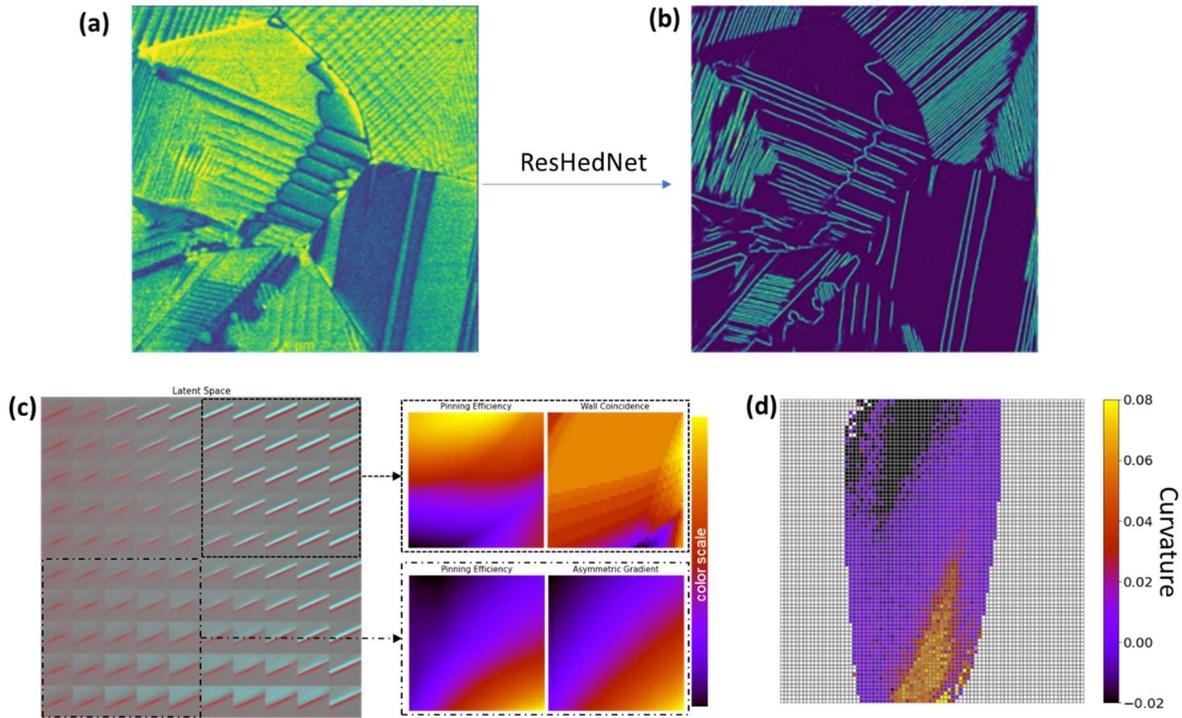

**Figure 6**. DCNN and VAE investigation of ferroelectric domains in scanning probe microscopy. (a) domains in a ferroelectric sample imaged by piezoresponse force microscopy, (b), ResHedNet converts the domain image to domain wall data. (c) VAE manifold2D representation shows that the ferroelectric domain walls pinning mechanism is related to ferroelastic domain walls distributions. (d) VAE manifold2D representation indicates the latent space is related to domain wall curvatures. Adapted with permission from Ref [135].

Similar insight can be derived from mesoscale SPM images. For example, it is the formation of domains that makes of ferroelectrics important for applications in sensors, actuators, capacitors, gates, and information technology devices.[142, 154-157] Therefore, exploring the mechanisms of the ferroelectric domain and domain wall dynamics has been a long-standing problem.[158-161] A ferroelectric domain is an area of aligned spontaneous electric polarization, usually an area in nanoscale or microscale levels. Piezoresponse Force Microscopy has emerged as a powerful tool for the visualization of ferroelectric domains and explore associated physical mechanisms. Domain walls—boundaries between domains with different polarization orientations—with the discontinuity of polarization often present complicated properties due to the presence of defects, disorder, and complex histories. In addition, owing to the polar nature of ferroelectric materials, the bound charge may result in charged walls, leading to strong coupling between domain wall and semiconducting properties.[162, 163] Therefore, domain wall behavior and switching mechanisms at the nanoscale level, including wall motion, domain nucleation, and pinning, root a broad spectrum of properties and applications of ferroelectric materials.

Shown in Figure 6 is an example of a ferroelectric domain and its dynamics imaged by piezoresponse force microscopy. The raw phase and amplitude are extracted from a video that indicates continuous domain evolution under applied bias. Here the domain walls can be identified by a Canny filter or deep convolutional neural network (DCNN) model, as shown in Figure 6c and 6d, which indicated segmented ferroelectric walls and ferroelastic walls, respectively. Then, the identified domain wall videos allow the exploration of relevant mechanisms by machine learning analysis. Indeed, in our earlier study, using rotationally invariant autoencoders (rVAE) we discovered factors affecting the ferroelectric wall dynamics.[164]

**g. Theory-in-the-loop.**

For atomic systems, one can run simulations and obtain result while the experiment is still running. For example, we can observe existing structure or assemble a structure in the STEM environment on the atomic level, but we can't really measure its electronic/magnetic properties (since it requires probing states at the Fermi level). So, we can use theory to get an idea of how 'interesting' or 'useful' these structures are and to help identify the next step (e.g., what shall we passivate our vacancies with to make sure they remain stable, etc.)

Using DL surrogate models, can in principle allow these behaviors in real-time, opening a connection to physical deep learning models. Physical models constructed using first-principles theory to quantum Monte Carlo (MC) and finite-element methods, spanning over quantum-mechanical to continuum scales, carry a multitude of information on structural, thermodynamic, and electronic properties of materials. Hence, introducing knowledge gathered from theoretical simulations can guide DL models to become more physics(science)-informed. Such understandings from simulations also help in decision-making process by underpinning causal hypothesis-driven[165, 166] mechanisms. The elaborative nature of theoretical models assists in establishing structure-property relations for which structured causal networks work well to answer associational interventional and counterfactual questions. Policies to draw inferences on effects of specific variables on functionalities of interest in a simulated environment can eventually be translated into active discovery frameworks for automated experiments. It is possible to even design active learning strategies that draw prior information[167] based on theoretical models for exploring both existing as well as unknown chemical spaces.

Be it the DL models, or causal networks or physics-informed optimization schemes for active learning or all combined, there are certain challenges to bringing 'theory-in-the-loop' applicable to all of frameworks. Although significant progress has been achieved over the years in high-performance computing, stark differences in time-, length scale and latency still exist between running an experiment versus performing one simulation representing the same scenario. For example, density-functional theory (DFT) and molecular dynamics (MD) simulations can optimally generate system sizes of Å to nanometer scales. The length of time required to perform one such simulation that is up to a few microseconds long can be multiple CPU hours, whereas STEM images are typically available at a fraction of a second. Semi-empirical methods such as density functional based tight binding (DFTB) and the extended tight binding method, enable simulations of large systems and relatively long timescales at a reasonable accuracy. These are considerably faster for typical DFT ab initio,[168] but still takes a few hours to converge. Defining regions of interest in an automated fashion followed by performing simulations on those 'patches' is a suitable alternative to draw inferences on a partial or entire system.[169-173] These can be achieved by multiple means, such as finding optimized fit-to-experiment structures via forward modeling and scalable data analyses.

Overall, frameworks bringing microscopic experiment and theory together in real time require capability for rapid exploration of features, analyses of images, quantification of associated uncertainties and most importantly, simulations performed in reasonable (comparable to experiment run time) timescale to establish an active feedback loop for guiding the next set of experiments.

**h. Rewards**

A key but often overlooked aspect of an automated experiment is the reward. Interestingly, on the domain side, it is often understood implicitly and incorporated both in the type of research and experiment planning. Basic science is associated with exploration, e.g., the discovery of new behaviors or finding generalized descriptions of the observed behaviors. Applied science targets exploitation, e.g., optimizing a specific aspect of materials functionality. However, even basic science has the exploration limited to a specific region of knowledge space, or constrained to a strong hypothesis, to make the scientific process tractable. The reward in this case is refining or falsifying the hypothesis. The combination of these elements gives rise to the current paradigm of hypothesis-driven science, allowing for serendipitous discovery.

Comparatively, the reward in ML is usually much more well-defined. For example, in computer games, it is a win. In automated driving it is to get from location A to location B in the least amount of time, with minimal use of fuel, and avoiding accidents and road rule-breaking. Hence, while multiple types of rewards, including external prescribed rewards, curiosity, or empowerment in reinforcement learning, and strategies to balance multiple rewards in multi-objective optimization, have been developed, the reward structure is usually straightforward and is derived externally or from simple information-theoretical criteria.

The microscopy experiments are associated with a much larger range of possible rewards, defined locally within the experimental context or more broadly with respect to the chosen project or experimental workflow. For example, for instrumentation and technique development, the reward is often the increase of spatial and energy resolution[174] or minimization of the beam damage. In fact, for the microscopy community, this is often the primary factor driving the development of the microscope over the technological cycle, or tuning the existing system during the experiment. At the same time, this reward is of less interest for domain communities, that aim to explore specific materials or quantify specific aspects of materials behaviors.

Implicitly, the reward structure drives the experiment led by a human operator, and the reward structure and time budget control the experiment planning. Correspondingly, microscopy offers an extensive playground for planning multi-level workflows, with resolution and stability optimization as a lower-level goal and the discovery and experiment planning as a higher-level goal. The former task is defined within a well-defined parameter space, for example the tuning of the scanning probe or electron microscope. At the same time, domain specific tasks offer an interesting scientific and sociological challenge. Can these models and workflows be made machine interpretable? Similarly, given these domain specific rewards, can ML algorithms discover new strategies for tuning the microscope to achieve them?

## 3. Current state of ML in microscopy

The first applications of modern deep learning techniques to scanning probe[175] and scanning transmission electron[176] microscopy were demonstrated in 2017 for the off-line data analysis. Since then, there has been a growing interest in using deep and machine learning to automate routine analysis of microscopy data (such as atom/defect/particle finding),[8, 177-181] learning symmetries,[186] and to extract physically meaningful latent parameters explaining high-dimensional observations.[135, 140, 141, 182-185] More recently, the pre-trained deep learning models were used to identify on-the-fly objects of interests, such as domain walls and atomic defects, in automated experiments in scanning probe and electron microscopy.

However, a significant and well-known limitation of deep learning models pre-trained on static datasets is that they often fail to generalize out-of-distribution (OOD), demonstrating poor performance (or even outright failure) when applied to data from outside the domain of training examples. The examples of OOD data in microscopy are images acquired under different acquisition parameters, presence of contaminants on the surface, sample damage due to e-beam irradiation, or change in the probe state due to unexpected interaction with a sample surface. While a combination of ensemble learning and iterative re-training partially addresses this issue for atom-resolved data[181], it doesn't solve the problem entirely. Furthermore, the approach based on pre-trained deep learning models assumes that we always know what structures are "interesting", while in many cases this is exactly what we want to discover.

One way to overcome the above limitations is to replace model(s) pre-trained on a static dataset with a model actively interacting with a data generation process. Recently, our group have used an *active* learning approach for learning a probabilistic structure-property relationship on-the-fly and using this 'knowledge' to sample next measurement locations.[187, 188] This approach is based on the idea that we often know what physical behavior (functionality) we are interested in and that it is encoded in spectra, and we want to identify (local) structural features where this behavior is maximized. We also want to achieve this with as few measurements as possible. The structure-property relationship of interest is then approximated at each step via a deep kernel learning model,[189] which is used to sample the next measurement locations expected to host the physical behavior of interest.

## 4. Way forward

The considerations in sections 2 and 3 suggest that there is a tremendous opportunity for research and development at the interface between machine learning and microscopy. While the potential benefits for the microscopy community *per se* are by now well realized and explored in multiple publications and opinion pieces,[190-192] the opportunities for ML community in microscopy domain are much less realized. Below, we summarize the classical components for the developments of an ML ecosystem, but aim to highlight the potential benefits and outcomes from ML side.

**a. Data repositories: useful but limited**

The initial success of ML research was based on (and is continuously measured against) the static datasets, the proverbial big data. While the definitions of big data evolved with time, it is now generally understood that it will refer to the data sets that is sufficient to sample the distribution from which the data is drawn, for example allow training of a generative model. Similar to many other fields, initial effort in microscopy was also dedicated towards the development of supervised models.[50, 193] By analogy with other fields, it has been proposed that large-scale data repositories can be universally useful. Some attempts to create these in microscopy have been made.[194, 195]

We argue that this may not be the case, at least not in the direct form. While the training examples are definitely useful, they cannot represent all the ways how the microscope can work. Even in STEM, the microscope variability can be very significant between different experiments, and the phenomena such as beam distortion due to the incomplete correction can be mixed with (as an example), the atomic column shape distortion due to octahedra tilts in the beam direction.[196, 197] For the simplest example, all ideal crystalline materials with the same lattice will look the same under microscope (down to lattice parameter). In SPM, the variability of the possible objects is even larger and magnetic domains, ferroelectric domains, polar regions in nanomaterials, or grains in polycrystalline materials can give rise to very similar contrasts.

While at the first glance counterintuitive, we argue that this is simply because the imaging methods are convolution of material and detection system. For the latter, detection system can have variabilities and modalities that far exceed that of the material variability (all crystalline materials look the same under the microscope). Comparatively, all vision tasks are predominantly based on human eye, or compared to the human eye baseline (e.g., in medical imaging analysis is verified via human labeled data sets/experts). Correspondingly, the value of the bespoke image databases can be minimal unless images are augmented by the detailed metadata on microscope parameters and sample composition and history.

At the same time, microscopy creates unique opportunities for the development of novel ML methods. For example, the image plus microscope metadata pairs offers ideal model systems for the invariant risk minimization problems due to the presence of strong inferential biases. In other words, unlike many other computer vision problems, it is well known that the answer exists and is unique. This further opens the pathway for developing ML methods that allows for problem-specific augmentation (variability of the imaging system), encapsulate and aim to discover data generation process, etc.

**b. Hyper language: tuning and manipulations**

On a fundamental level, every microscopy experiment consists of a carefully orchestrated sequence of tasks, often with inner and outer optimization loops, and decision making at particular steps. This appears obvious when stated but is in fact so ingrained in the standard microscope operator's daily routine that it is rarely formulated in explicit form. However, in order to enable fully autonomous microscopes, we need to develop a language that can represent every low-level task that a microscope system can perform. This can be thought of as the microscopy equivalent of the recently developed 'XDL' programming language for chemical synthesis, which is a Turing-complete language that is in principle able to represent all types of chemical reactions.[198]

Naturally, the low-level microscopy functions would be operations such as moving the stage and/or tip to a certain location, scanning along a trajectory, enabling and disabling feedback systems, measuring along different channels, changing local and global excitation parameters, and so forth, and there would additionally be 'decision blocks' for where decisions would be made about the next course of action (either by a human or algorithm).

As an example, consider a simply scanning probe microscopy experiment of measuring the roughness of a sample. The basic steps would be to initiate the tapping mode operation (including the tuning steps), scan regions of a particular size that are free from large debris, and move the scan window after each scan to collect more images and thus statistics. Then the tip should be withdrawn, and the images analyzed with a standard topographic analysis to determine the average sample roughness. This could easily be represented as a sequence of function calls in a script, but it could also be represented by an even more compact representation akin to that of molecules – such as the SELFIES representation introduced by Aspuru-Guzik group.[199, 200] Then, performing an experiment and generating new experiments becomes an exercise in generating new such 'microscope SELFIES'.

**c. Beyond human workflows**

An interesting and very weakly explored concept in the context of automated experiment is the development of beyond-human workflows. Indeed, currently most of the effort in automated experiment relies on classical human based workflows, implemented via robotic agents. An alternative is combined human and robotics operators orchestrated by central agent.

At the same time, recent advances in reinforcement learning have demonstrated that for closed systems with known rules but complex emergent strategies (e.g. Alpha Go), AI is capable of formulating the strategies beyond human. Interestingly, this leads to the inverse developments when exposure to the AI agents improves human learning. Until now, these developments have been limited to fully closed system, albeit recent progress with large models such as DALLE make in the art field suggest that the situation is likely to change,

Similar developments, albeit with a certain time delay, are now proceeding in the field of automated labs. To date, the vast majority of automated or hybrid labs utilize the human made workflows derived from human experts or via literature mining.[201-205] In many cases, ML is used to optimize specific steps such as yield of specific product via optimization in low-dimensional continuous parameter space and implemented via continuous flow reactors, microfluidic systems, or pipetting robots.[206] However, proponents such as Lee Cronin[207] now explore pathways for full chemistry design with 3D printing of components. That said, exploration of the chemical space of e.g. small molecules is an exceptionally complex problem due to its immense size and non-differentiability.[208] For the experimental world, this often limits discovery space to the molecules that can be synthesized from a limited set of precursors and reactions.

Comparatively, microscopy offers a large, but still much smaller realm for exploration. Systems such as atomic manipulation by electron beam or STM probe or ferroelectric domain manipulation by PFM offer the richness of unknown physical behaviors, a multitude of possible metastable states, and a variety of ways to reach them. However, these systems come with a finite range of possible actions and actions that are fully available within the specific system. As such, this can be an ideal toy system for exploring ML creation of workflows starting from optimization of human based ones but evolving to discover new ones, via empowerment learning or other techniques. Interestingly, even Deep Kernel Learning discovery of the plasmonic EELS signals already presents beyond human workflow.

### d. Learning physics

Generally, a physics-based solution, in circumstance where the physics is computable and understood, is much preferable to a purely data-driven approach, due to efficiency of learning. In most cases, the physics-models may have zero, one or a few parameters to be estimated. As an example, consider that that the standard 'cartpole' reinforcement learning environment, where the aim is to move the cart left or right to balance a pole in the center of the cart without it tipping over, is solvable using deep reinforcement learning techniques, which can take 100s of episodes, or, it can be solved in 5 lines of code without any RL, given knowledge of Newtonian mechanics.[209] Of course, in a real scientific problem, we may have to infer the underlying model from noisy measurements of the system, often sparsely in space and/or time.

Notably, the ML community is developing methods that can attempt to recover the underlying equations that generate the observations measured, through techniques such as symbolic regression. These are exemplified by the approach of the 'AI Feynman' proposed by Tegmark,[210-212] amongst others.[213, 214] In general, the applications of these techniques to microscopy are still very limited. For instance, it would be instructive to learn the force-fields from an atomically resolved STEM video of an amorphous material being crystallized by the electron beam, or of directly determining the energy of barriers for dopants imaged moving through a 3D lattice. Although much simpler in the static case,[215] the extension to a fully dynamic system naturally tends to lead towards the need for large volumes of data, and careful tuning of thermodynamic variables to efficiently determine the likelihood of proposed models describing the dynamics.

### e. Causal processes, interventions, and counterfactuals

Traditional causal discovery approaches assume that the observations are structured into elementary units representing random variables connected by a causal graph. In some cases, we may treat different modalities in multimodal scanning probe microscopy as such units. Practically, an image in each observational channel is split into patches at selected coordinates, followed by an established post-processing analysis to form appropriate structural and functional descriptors associated with such units. Then, one can deploy causal discovery models for finding causal links between various structural and functional parameters.

At the same time, in many experiments, the observations cannot be directly mapped onto a causal graph, and one must first uncover ("learn") in an unsupervised manner low-dimensional, high-level causal variables from high-dimensional, low-level observational data to allow for causal discovery. The emerging field of causal representation learning[216] aims at addressing this problem through an extension of the deep latent variables (correlative) models. The goal of deep latent variable models, such as (variational) autoencoder ((V)AE), is to explain high-dimensional observations by a small number of unobserved (latent) variables.[217] Recently, VAEs were used to uncover physical order parameters in lattice models and dynamic atomically resolved imaging data.[183] However, the latent variables in those VAE setups did not support intervention and reasoning. Recently, an approach based on self-supervised learning with data augmentation has been proposed to overcome this limitation of traditional deep latent variable models. In this approach, data augmentation acts as a counterfactual under a hypothetical intervention, allowing to separate content from style. We argue that in microscopy, one could replace data augmentations by changing various imaging parameters during the experiment to disentangle the real physical variables and use them for further causal discovery.

## 5. Future

Finally, we aim to summarize what these developments can bring to the classical physics, chemistry, and materials science fields, including both the extant ones, as well as new areas that may emerge based on these developments. While the list below is undoubtedly partial and new applications can be expected to be ideated and implemented, it nonetheless suggests an extremely broad potential for emerging science at the interface between machine learning and microscopy.

### a. Learn physics

Traditional physics describes the structure of crystalline materials as ordering of building blocks (elementary unit cells) in a periodic lattice. These descriptors are conveniently matched to the information that can be obtained from scattering methods, and in turn allows exploring materials structure in the Fourier domain. However, many important materials such as highly defective solids, structural, spin, and cluster glasses, and ferroelectric relaxors do not have periodic structures.[218, 219] While their structures can be visualized via electron microscopy, it is not clear what information can be derived from the knowledge of the atomic positions. Similar opportunities arise in the context of the electron microscopy of the extended defects and multiphase materials, scanning tunneling microscopy of materials with complex topological and electronic orders, and even conventional SPM of ferroelectric and ferromagnetic domains and materials microstructures. The microscopy images are often spectacular, contain information on the specific aspects of materials structures and functionality, but extracting this information is non-trivial.

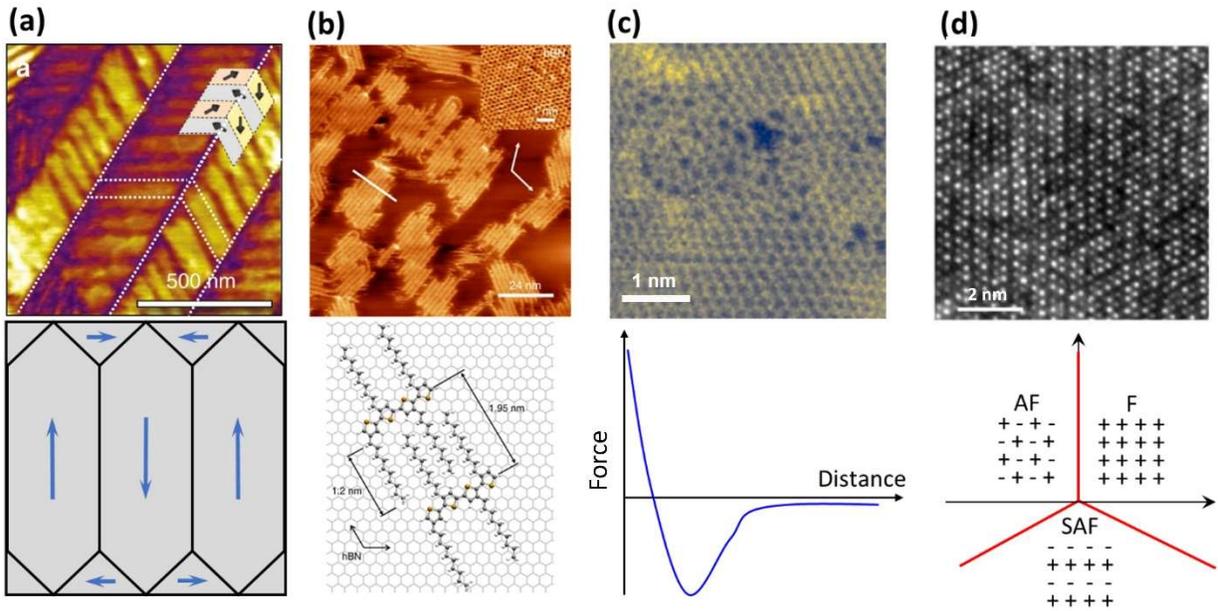

**Figure 7.** (a) Force-SPM phase image shows domain structure in $PbZr_{0.2}Ti_{0.8}O_3$. Adapted with permission from ref[220] (b) Force-SPM images of P3DT adsorbed on the surface of hBN. Adapted with permission from ref[221] (c) STEM image of disordered graphene, , and force-distance curve. (d) STEM of a $MoS_2$ doped with Re atoms, adapted with permission from Ref[222] with a snapshot of the Ising model.

One approach to use these data is to train the generative statistical models such as VAEs and GANs, that can provide multiple examples of the microstructure from a smaller number of observations. However, such models typically extrapolate very poorly, meaning that the data acquired for specific set of conditions will not generalize to a different one. At the same time, the structural complexity of many materials strongly depends on temperature. For example, in materials undergoing order-disorder phase transition, correlation length (and hence the size of structural descriptor) diverges in the vicinity of phase transitions

Comparatively, the complexity of observed structures and functionalities of solids emerges from relatively simple interactions. For atomic distributions in solid solutions and certain magnetic systems, these can be the variants of an Ising Hamiltonian, representing the interactions in the system via pairwise exchange integrals. For structurally amorphous systems, the system can be described via force fields acting between the atoms. Note that the modelling of the systems behavior from known force fields or Hamiltonians is the foundational part of computational physics. However, relevant materials parameters are usually derived *ad hoc*, or fit from a large body of macroscopic observations.

Hence, the obvious challenge for the ML in physics is probabilistic learning of the atomic interactions on the level of lattice models, force fields from the observational data, or phase field free energies. These physical generative models can provide insight into the fundamental physics and chemistry of materials.

### b. New chemistry

Synthetic organic chemistry is a cornerstone of fields ranging from drug and catalyst design to polymers. Applications of machine learning in chemical synthesis to understand the relationships between molecule structure and biological functionality, predict the synthesizability, and predict possible synthetic pathways from available components is one of the most rapidly developing scientific directions,[223, 224] via graph networks,[5, 225] etc. The rapid emergence of cloud laboratories provides a real-world implementation for *in-silico* developments. However, by its very nature chemical space is extremely high-dimensional, and exploring even specific classes of compounds often requires access to large number of initial compounds and ligands.

The emergence of imaging methods allows us to merge chemistry and characterization. The first definitive example of this synergy is the overlap between CryoEM and biology, where local structural information on biomolecules is now being used to design biological targets, etc. Electron microscopy methods can provide similar insights into the chemistry of solids. For example, STEM data can be used derive a library of possible defects in solids,[177] We can also extend this to explore whether we can chemically modify them and build reaction schemes on top of it as a pathway towards better catalysts, etc.

### c. Nanophotonics and quantum optics

The wavelength of light, about half a micron, fundamentally limits its use in nanotechnology – hence the primary adoption of electrical signals to this day. Currently, using extreme ultraviolet (EUV) exposure methods, the semiconductor industry has pushed the limits of creating nanoelectronic devices down to the 3~nm node length, but such devices all rely on the use of electrons or holes for computations and information transfer, which inefficiently generate heat and is relatively slow. On the other hand, light propagates at the ultimate speed (of light!) and because photons do not have a rest mass, it does not lose energy to Joule heating. However, it cannot be easily confined to volumes smaller than the diffraction limit, or about 250 nm.

Plasmonics offers a tempting route to bridge the macro and nano scales with light by exploiting a quasiparticle known as the plasmon, which is light coupled to a collective electron wave at the interface of (usually) a metal and insulator, either propagating across an interface or localized to a small volume. The former is known as a surface plasmon polariton (SPP), while the latter is known as a localized surface plasmon resonance (LSPR). Intriguingly, a strong enhancement of the light can occur when it is coupled to the surface plasmon,[226] making for a useful attribute. An idea for information transfer is that an incoming light source on the order of millimeters in size is focused down by ordinary means close to the diffraction limit and then partially transformed into a propagating wave (SPP) that is tens of nanometers in size, therefore the nature of the plasmon acts as a bridge between these two different length scales. Other applications take advantage of the LSPR in order to ensure extremely localized (and enhanced) light or heat sources. For example, in some cancer treatments, small nanoparticles may be placed in the body and then illuminated by a light source (to which the human body is transparent), which in turn excite LSPRs in the nanoparticles that act as tiny, localized heat sources to attack cancerous cells.[227]

In terms of quantum optics, controlling the location, energy, and spatial extent of an emitter is of utmost importance, particularly in the rapidly developing field of quantum computing. Plasmonic nanoparticles can act as localized emitters that can in turn excite, for instance, a dye molecule which itself may be a single photon emitter. Alternatively, and perhaps of more utility, point defects in 2D materials like boron nitride (BN)[228] can also act as quantum emitters, and the electron beam can potentially be used to both create and probe such single defects.

The STEM then offers a unique and well-suited platform for nanophotonics.[229] From the point of view of measuring the light-matter interaction's energy dependence in space at the nanometer scale, STEM-EELS is the only currently available tool to spatially resolve the energy dependence of plasmon modes. A perfect corollary to the nano and atomic scale measurement capability is the ability of the STEM to geometrically sculpt or chemically modify a two- or three-dimensional solid such that it suits a specific plasmonic need, which can then be measured in STEM-EELS almost immediately.

**d. Atomic fabrication**

Fabricating matter atom by atom has long remained the dream of scientific community. In his final note, Richard Feynman had famously written "what I cannot create, I do not understand".[230] It is clear that atomic fabrication[231-234] at scale is a pathway for quantum devices, new electronics, and serendipitous discoveries in multiple areas of science and technology.

Currently, both STM and STEM can be used to manipulate matter on the atomic scale, as have been demonstrated by multiple groups over time.[65, 235, 236] However, while the enabling instrumentation and the engineering controls are currently available, the operation of the microscopes is preponderantly enabled by human operator. This is a clear area where machine learning can make a tangible difference. For STM, the relevant tasks include tip conditioning and optimization, rapid sampling of surface electronic structure, investigating the electronic properties of individual atomic groups, physics discovery based on desired electronic signatures, and learning the rules of atomic manipulation towards specific functionalities and structures. Similar problems emerge in the STEM context, with added complexity of the higher stochasticity of the system.

## 6. Our vision

The last decade has seen the exponential growth of machine learning methods that by now have become an inseparable part of multiple scientific and applied domains. However, the emerging challenge in ML is a transition away from static data sets towards exploring active data generation processes and enabling autonomous systems, from automated cars to chemical design and synthesis. Each of these fields encounter problems related to the inability of classical ML to generalize and extrapolate. At the same time, many of these fields own specific physical interferential biases, are many hidden degrees of freedom, as well as exogenous variables.

We argue that microscopy platforms that operate in very simplified (compared to the macroscopic world) physical environments, and that are associated with strong physical biases and low risks, are ideal toy system for deployment of active learning algorithms, learning physics based digital twins, and many other applications. While these tools have traditionally been perceived to be highly specialized, over the last two decades microscopes have evolved to become significantly more user-friendly, with effective APIs and easy to use systems capable of visualizing matter from atomic to micron levels. Further, they are capable of a broad spectrum of actions, ranging from manipulation of nanoparticles and nanowires, to inducing local chemical reactions, and moving single atoms and making and breaking chemical bonds. From the domain perspective, the missing component is the algorithms that enabled controlled manipulation towards specific materials objectives. At the same time, from the ML perspective, these domains offer a challenging but bounded environment with characteristics ideal for development and deployment of active learning and reinforcement learning methods. In contrast to many other traditional environments (e.g., for robotics), these systems are associated with a reduced set of commands, low risk, and well understood physical biases and causal relationships, often with minimal exogenenous effects.[237] In other words, microscopy is all you need (to start).

**Acknowledgements:**

This research is sponsored by the INTERSECT Initiative as part of the Laboratory Directed Research and Development Program of Oak Ridge National Laboratory, managed by UT-Battelle, LLC, for the US Department of Energy under contract DE-AC05-00OR22725. This effort was primarily supported by the center for 3D Ferroelectric Microelectronics (3DFeM), an Energy Frontier Research Center funded by the U.S. Department of Energy (DOE), Office of Science,

Basic Energy Sciences under Award Number DE-SC0021118. This research was supported by the Center for Nanophase Materials Sciences (CNMS), which is a US Department of Energy, Office of Science User Facility at Oak Ridge National Laboratory. The authors are grateful to Prof. J. Agar (Drexel) and K. Brown (Boston University) for early comments on the manuscript.